\documentclass[sigconf]{acmart}

\usepackage{amsfonts}
\usepackage{bm}
\usepackage[linesnumbered,ruled,vlined]{algorithm2e}
\usepackage{pifont}
\usepackage{float}
\usepackage{graphicx}
\usepackage{makecell} 
\usepackage{wrapfig}

\copyrightyear{2024}
\acmYear{2024}
\setcopyright{acmlicensed}\acmConference[ICCAD '24]{IEEE/ACM International Conference on Computer-Aided Design}{October 27--31, 2024}{New York, NY, USA}
\acmBooktitle{IEEE/ACM International Conference on Computer-Aided Design (ICCAD '24), October 27--31, 2024, New York, NY, USA}
\acmDOI{10.1145/3676536.3676714}
\acmISBN{979-8-4007-1077-3/24/10}

\begin{document}

\title{Voxel-CIM: An Efficient Compute-in-Memory Accelerator for Voxel-based 
Point Cloud Neural Networks}

\author{Xipeng LIN}
\affiliation{%
  \institution{The Hong Kong University of Science and Technology(Guangzhou)}
  \city{Guangzhou}
  \country{China}
}
\email{xlin297@connect.hkust-gz.edu.cn}

\author{Shanshi Huang}
\affiliation{%
  \institution{The Hong Kong University of Science and Technology(Guangzhou)}
  \city{Guangzhou}
  \country{China}}
\email{shanshihuang@hkust-gz.edu.cn}

\author{Hongwu Jiang}
\affiliation{%
  \institution{The Hong Kong University of Science and Technology(Guangzhou)}
  \city{Guangzhou}
  \country{China}
}
\email{hongwujiang@hkust-gz.edu.cn}

\renewcommand{\shortauthors}{Lin et al.}


\begin{abstract}

  The 3D point cloud perception has emerged as a fundamental role 
  for a wide range of applications. In particular, with the rapid development of neural networks, the voxel-based networks 
  attract great attention due to their excellent performance. Various accelerator 
  designs have been proposed to improve the hardware performance of voxel-based networks, 
  especially to speed up the map search process. However, several challenges still exist 
  including: (1) massive off-chip data access volume caused by map search operations, 
  notably for high resolution and dense distribution cases, (2) frequent data movement for data-intensive 
  convolution operations, (3) imbalanced workload caused by irregular sparsity of point data. 

  To address the above challenges, we propose Voxel-CIM, an efficient  Compute-in-Memory based 
  accelerator for voxel-based neural network processing. To reduce off-chip memory access for map search,
  a depth-encoding-based output major search approach is introduced to 
  maximize data reuse, achieving stable $O(N)$-level data access volume in various 
  situations. Voxel-CIM also employs the in-memory computing paradigm and designs 
  innovative weight mapping strategies to
  efficiently process Sparse 3D convolutions and 2D convolutions. 
  Implemented on 22 nm technology and 
  evaluated on representative benchmarks, the Voxel-CIM achieves averagely 4.5\~{}7.0$\times$ higher 
  energy efficiency (10.8 TOPS/w), and 2.4\~{}5.4$\times$ speed up in detection task 
  and 1.2\~{}8.1$\times$ 
  speed up in segmentation task compared to the state-of-the-art point cloud accelerators 
  and powerful GPUs.

\end{abstract}



\maketitle

\section{Introduction}
\label{sec:introduction}
The 3D point cloud is a mass of spatial points
obtained from a variety of 3D sensors, such as LiDAR, cameras, etc., which serves as 
a fundamental role
in various applications, ranging from VR/AR to autonomous vehicles and robotics.
Unlike 2D images, the inherent complexity of point cloud data, characterized by its random 
and irregular spatial sparsity, poses huge challenges for point cloud processing algorithms.
\begin{figure}[t]
  \centering
  \includegraphics[width=\linewidth]{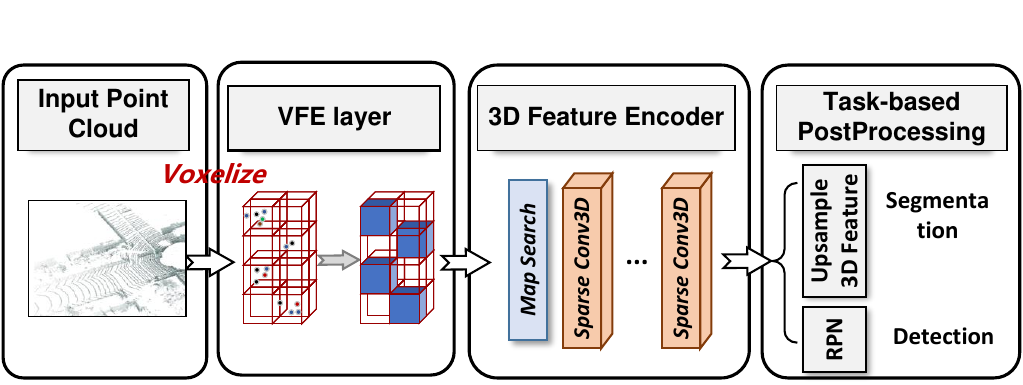}
  \caption{The generic network architecture of voxel-based algorithms for segmentation
  and detection tasks.}
  \label{fig:figure1}
  \Description{}
\end{figure}

In recent years, a variety of algorithms based on neural networks have been proposed to 
address these challenges \cite{qi2017pointnet, qi2017pointnet++,  yin2021center}. 
Among them, the voxel-based network shows excellent 
performance \cite{zhou2018voxelnet, yan2018second, lang2019pointpillars, shi2020pv}. Instead of handling 
huge amounts of irregular point data directly,
the voxel-based network chooses to partition the continuous 3D space into volumetric 
pixels and 
voxelizes point data into structured formats. As shown in Fig. ~\ref{fig:figure1}, a typical voxel-based network
consists of three stages:
voxel feature extractor (VFE), 3D Feature Encoder, 
and task-based postprocessing network. VFE layer is responsible for encoding the features of each voxel.
The 3D Feature Encoder is composed of stacked Sparse 3D convolution (Spconv3D) operations.
To facilitate Spconv3D, map search is conducted to avoid convolutions on empty voxels.
For task-based postprocessing network,
a region proposal network (RPN) is generally adopted to generate detection results
(e.g., SECOND \cite{yan2018second}) in detection tasks. And 
the Unet structure is usually
used (e.g., MinkUnet \cite{choy20194d}) in segmentation tasks. 
Despite showing excellent accuracy performance, the throughput of voxel-based networks is usually limited on traditional computing platforms such as GPU. 
(e.g. SECOND runs 36 FPS  
 on Nvidia 3090ti and MinkUnet runs 13 FPS on Nvidia 2080ti \cite{lyu2023spocta}). 
 
 Although many efforts have been made to accelerate point cloud neural networks,
There are still three main challenges that restrict the  
hardware performance of voxel-based networks.
\textbf{1) The mapping operation is dominated by huge off-chip data access.} Map search, a classic operation prior to Spconv3D, aims 
to build IN-OUT maps to accelerate following 
Spconv3D. Previous acceleration methods can primarily 
be divided into two categories, table-aided and table-free map search. Table-aided strategies 
used hash tables or octree-encoding-based tables, where all voxels 
are encoded \cite{choy20194d,tang2020searching,tang2022torchsparse,cao2022sparse,lyu2023spocta}.
Although table-aided methods can achieve $O(1)$-level searching speed theoretically, 
The table requires a large storage capacity, potentially 
exceeding 100MB. On the other hand, The table-free strategies transform the search problem
into an intersection-detecting problem, avoiding the huge table
\cite{lin2021pointacc, yang2023efficient}.
However, this method introduces redundant off-chip memory access for input data.
\textbf{2) Frequent data transfers for data-intensive computations.}
In traditional Von Neumann architecture, the "memory wall" problem limits the processing speed of neural network calculations since
massive data movement happens between the computing units and
the storage units, resulting in low throughput \cite{yu2021compute}. Especially, both
activations and weights need to be fetched from the storage unit.
Moreover, most power consumed in conventional ASIC accelerators
for DNNs is also from memory access operations.
\textbf{3) Imbalanced workload caused by irregular sparsity of point data} \cite{tang2022torchsparse, hong2023exploiting}.
Due to the sparsity, randomness, and uneven distribution of point clouds, 
each weight corresponds to a different number of in-out pairs. 
Consequently, the computational workload is usually high for central weights while keeps
low for weights on the edge. The gap between the workload of the central weights and the
peripheral weights could be huge, leading to low computational
resource utilization.

To address the above challenges, we propose Voxel-CIM, 
an efficient Compute-in-Memory (CIM) based accelerator for voxel-based neural network
processing with a novel map search method.
Our key contributions can be summarized as follows:
\begin{itemize}
  \item To reduce the off-chip data access volume, we proposed a 
  novel searching scheme, 
  named by Depth-encoding-based Output Major Search (DOMS),
  which combines the output major searching approach and depth-encoding table. DOMS can 
  maximize data reuse, achieving stable $O(N)$-level off-chip memory access in various situations.
\end{itemize}
  
\begin{itemize}
  \item 
  We design a CIM processing unit and corresponding weight mapping strategies to 
  process Spconv3D and Conv2D calculations efficiently. We also propose a
  weight workload balanced (W2B) method to solve the workload mismatch problem,
  which can realize a 2.3$\times$ speedup in the segmentation benchmark.

\end{itemize}

\begin{itemize}
    \item 
    We comprehensively evaluate the performance of our proposed accelerator 
    on detection and segmentation benchmarks, showing
  that our accelerator could achieve 
  averagely 4.5\~{}7.0$\times$
higher energy efficiency (10.8 TOPS/w), and 2.4\~{}5.4$\times$ speedup
in detection task and 1.2\~{}8.1$\times$ speed up in segmentation task 
compared to the state-of-the-art point cloud accelerators and powerful GPUs.

\end{itemize}

\section{Background}

\vspace*{7pt}

{\noindent\itshape A. Map Search Operation}


Map search is a critical operation to accelerate subsequent sparse convolution by 
building the IN-OUT maps. 
We define 
the input and output voxel points as $P_i$ and $Q_j$, where $P_i$, $Q_j$ $\in \mathbb{Z}^3$
are the quantized coordinates in Z-dimensional 
space. 
For each output point ${Q_j}$, we need to traverse all its adjacent 
locations with kernel offsets  $\Delta^{3}(K)$ (K is the size of kernel, e.g., $\Delta^{3}(3) = \{-1,0,1\}^{3}$).
Each valid pairs of input 
point ${P_i}$ and output point ${Q_j}$ constitutes one item in the
IN-OUT maps, defined as $\mathcal{M}(j) = \{({P_i}, {Q_j}, W_{\delta})|\delta \in \Delta^{3}(K)\}$. This map will guide 
subsequent sparse convolutions, thereby avoiding performing convolutions on
empty voxels \cite{graham20183d}.

\vspace*{7pt}

{\noindent\itshape B. Sparse 3D Convolution}


A sparse tensor representation consists of two parts: the voxels' spatial
coordinates and feature vectors.
Mathematically, the collection of sparse tensors $T$ of one layer can be represented as:

\begin{equation}
  P = \begin{bmatrix}
    x_1 & y_1 & z_1 \\
    \vdots & \vdots & \vdots\\
    x_N & y_N & z_N
    \end{bmatrix} P_i \in \mathbb{Z}^3, \quad F = \begin{bmatrix}
    f_1^T \\
    \vdots \\
    f_N^T
    \end{bmatrix}f_i \in \mathbb{R}^C
\end{equation}
where \( P_i = (x_i, y_i, z_i) \) is the coordinates of the non-zero voxel, 
\( f_i \) is the associated feature vector, and \( N \) is the number of non-zero voxels.
For a sparse 3D convolution with a kernel size \( K \) applied to a sparse 
tensor \( T \), the output feature at coordinate \( Q_o \) is:
\begin{equation}
f'_o \ = \sum_{\delta \in \Delta^{3}(K)} W_{\delta} f_i,\ \  for (P_i, Q_o, W_{\delta}) \in \mathcal{M}(o)
\end{equation}
where \( f'_o \) is the output feature at coordinate \( Q_o \), 
 and \( \mathcal{M}(o) \) represents all in-out pairs associated to the output coordinate \( Q_o \).

There are three different kinds of Spconv3D as follows:

\begin{itemize}
  \item {\itshape Submanifold Spconv}. The submanifold spconv preserves the 
  spatial location of voxel data, and is typically used for spatial feature 
  embedding in the middle of a block. 
  The common parameter setting includes a kernel size of 3 and a stride of 1, 
  denoted as subm3.
\end{itemize}

\begin{itemize}
  \item {\itshape Generalized Spconv}. The generalized spconv is often 
  positioned at the end of a block as 
  a downsampling layer. It operates under a scheme where an output is 
  considered valid if any inputs fall within the kernel range. Consequently, 
  the output tends to be more dispersed in spatial distribution. 
  The typical parameter setting includes a kernel size of 2 and a stride of 2, 
  denoted as gconv2.
\end{itemize}
\begin{itemize}
  \item {\itshape Transposed Spconv.} The transposed spconv is 
  essentially the reverse process of the generalized spconv and is 
  commonly employed for upsampling in segmentation tasks. It follows the same 
  computational rules as the generalized spconv.
\end{itemize}

\vspace*{7pt}

{\noindent\itshape C. Region Proposal Network}


RPN is widely used in detection 
frameworks \cite{girshick2015fast, lin2017focal}, which deals with 
dense and structured tensor features to generate the final detection
results.
Typically integrated with a pyramid structure \cite{lin2017feature}, 
an RPN comprises three blocks of stacked 2D convolution (Conv2D) layers.
Each block downsamples the feature map with a stride of 2. 
The last two blocks upsample their features and concatenate them with those of first
block. Then the concatenated feature is used for subsequent detection.

\begin{figure*}[t]
  \centering
  \includegraphics[width=\textwidth]{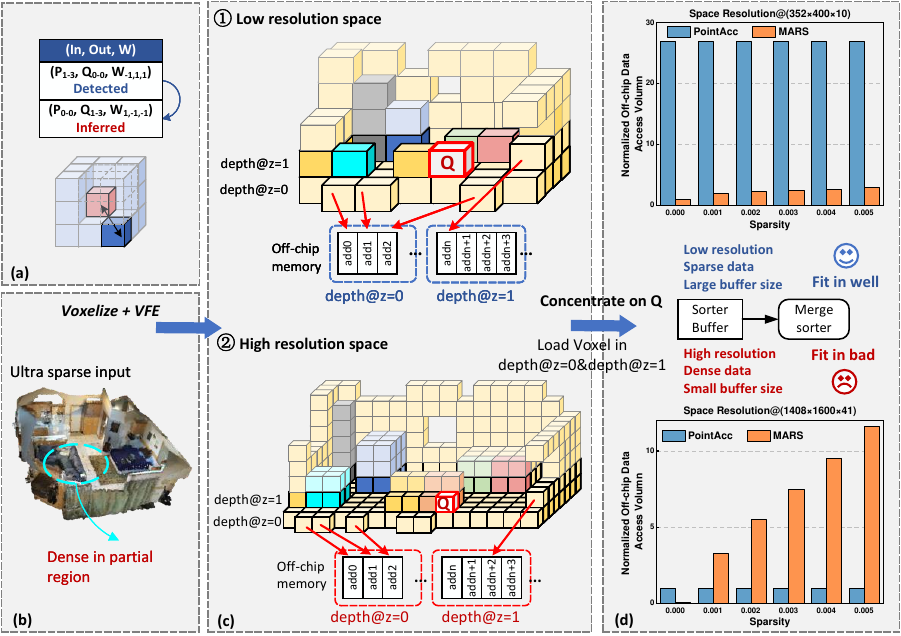}
  \caption{(a) The reverse mapping pairs can be inferred due to symmetry.
  (b) The visualization of input cloud data.
  (c) Voxelization with low resolution VS. voxelization with high resolution. 
  (d) The comparison of normalized off-chip data access volume in various situations.
  (To simulate buffer limitations in extreme cases, we set
  the buffer size to match the length of the merger sorter, which is 64.)}
  
  \label{fig:mars}
  \Description{}
\end{figure*}

\section{VOXEL-CIM accelerator design}

\subsection{Effective Depth-encoding-based Output Major Searching Method}

\vspace*{7pt}

{\noindent\itshape A. Previous Methods}


 Previous table-free map search methods can be divided into two categories, weight-major approach and output-major approach.
 
 \textit{Weight-major Approach.} 
 This approach is proposed by 
  PointAcc \cite{lin2021pointacc}.
  It iterates and loads all voxels for each weight to find intersection pairs.
  The on-chip buffer is not large enough to hold all voxels at a time, which causes
  redundant memory access. If it has $N$ voxels and the 3D convolution kernel with size $K$, 
  the off-chip memory data access volume can reach up to $O(K^3 \times N)$.

  \textit{Output-major Approach.} 
  This approach is proposed by MARS \cite{yang2023efficient}. 
  It concentrates on each output voxel, 
  maximizes voxel reuse, and achieves $O(N)$ for off-chip memory access theoretically. 
  It also utilizes the symmetry of the conv3D kernel to reduce half of the 
  mapping calculation. As shown in Fig.~\ref{fig:mars}(a), the $3 \times 3 \times 3$
  kernel is central symmetry. If pair $(P_{1-3}, Q_{0-0}, W_{-1,1,1})$ exists, 
  a reserve mapping pair $(P_{0-0}, Q_{1-3}, W_{1,-1,-1})$ must exist. This implies that
  instead of searching the 26 surrounding positions, it is sufficient
  to examine half adjacent positions, merely 13. 

To ensure that exhaustively search all possible in-out pairs 
for each output with a single load, it must be able to accommodate 
two depths' voxels at least, which raises the size requirement for sorter buffer.
As depicted in Fig.~\ref{fig:mars}(c) \textbf{\ding{172}}, 
for the output voxel Q, the searching space is restricted to two depths 
(depth@z=0 and depth@z=1, marked in darker color).
In most cases, the voxel space is ultra sparse, and the sorter buffer is 
sufficient to hold the voxels of two depths.
However, there can be dense distributions in some partial regions, 
as shown in Fig.~\ref{fig:mars}(b). 
What's worse, with the introduction of simpleVFE \cite{GithubRepo} and the application 
of complex scenarios, 
the network tends to perform Spconv3D on high-resolution voxel spaces.
As illustrated in Fig.~\ref{fig:mars}(c) \textbf{\ding{173}}, voxelization with high resolution will 
generate large scale voxel
space to keep detailed surrounding information.
If the sorter buffer is not enough to store two depths' voxels, 
this method will require multiple loading
to exhaustively search all space for one output.
As illustrated in Fig.~\ref{fig:mars}(d), when the search space is low resolution,
the data distribution is very sparse, and the sorter buffer is large enough, MARS can realize 
optimal $O(N)$ access.
If the search space is high resolution and the data distribution is dense relatively,
the data access volume deteriorates rapidly due to buffer limitation and
repeated loading.

\vspace*{7pt}

{\noindent\itshape B. Insight}

\begin{figure}[t]
  \centering
  \includegraphics[width=\linewidth]{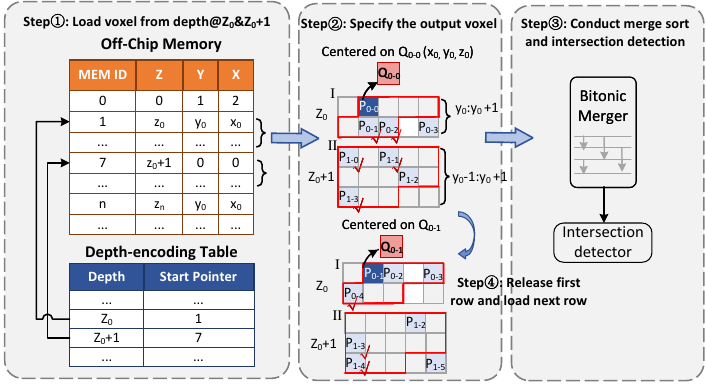}
  \caption{The workflow of DOMS
    }
  \label{fig:doms}
  \Description{}  
\end{figure}

The essential problem is that voxels' size of two depths 
is still too large to fit in the sorter buffer and merge sorter circuit in
extreme cases.
Actually, there is no need to accommodate voxels of two continuous depths. As shown in 
Fig. ~\ref{fig:doms}, 
for an output voxel $Q_{0-0}(x_{0}, y_{0}, z_{0})$, the valid searching space is restricted to two 
consecutive rows $(:, y_{0}:y_{0}+1, z_{0})$ at the same depth and three 
consecutive rows $(:, y_{0}-1:y_{0}+1, z_{0}+1)$ at the next adjacent depth.
If the position of start voxel of each depth can be located in off-chip memory,
we can load the voxels of corresponding depth directly.
Thus, we can build a depth-encoding table to
record the start pointer of each depth in off-chip memory. For each 
output voxel, we can locate the start position of corresponding depth, and tile two rows of the same depth 
and three rows of the next depth to ensure complete searching space.

\vspace*{7pt}

{\noindent\itshape C. Depth-encoding-based Output Major Search}


Fig. ~\ref{fig:doms} illustrates the tiling and scheduling strategy.
Assuming that each depth's buffer can hold only 4 voxels, 
the search process is as follows:
\textbf{Step\ding{172}}: In the beginning,
the voxels of first two depths $(z_0\, and\, z_0+1)$ are tiled to 
corresponding buffer $(I\, and\, II)$ respectively as many as possible;
\textbf{Step\ding{173}} Specify the output voxel and calculate its 13 adjacent positions by adding
kernel offset; \textbf{Step\ding{174}} Pack two depths' voxel and output adjacent positions to merge
sorter, and detect intersection; \textbf{Step\ding{175}} Then, the data of first row of two buffer
$(I:P_{0-0}@z_0\, and\, II:P_{1-0}, P_{1-1} @z_0+1)$ will be released and the data of 
margin rows $(I: P_{0-1}, P_{0-2}, P_{0-3}@z_0\, and \,II: P_{1-2}, P_{1-3} @z0+1)$ will
be reused for next search. For the output of next row, we need to load data of next
row $(I: P_{0-4}@z_0\, and \,II: P_{1-4}, P_{1-5} @z_0+1)$ for processing. 
Keep processing until the entire depth is fully handled.
After completing the search of the depth$@z0$, the target output moves to depth$@z0+1$. 
The buffer I will be cleared, and new voxels in depth$@z0+2$ will be loaded.

\begin{figure}[t]
  \centering
  \includegraphics[width=\linewidth]{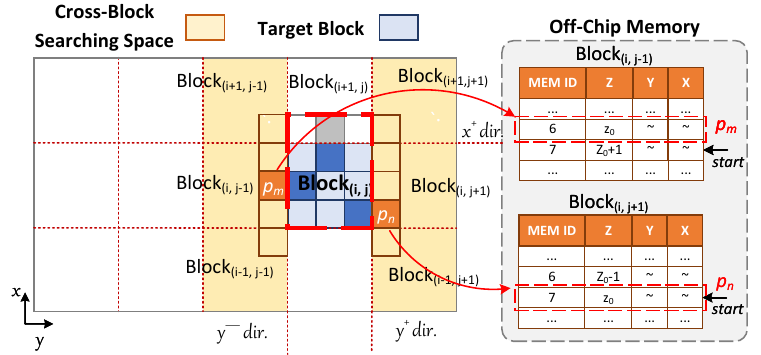}
  \caption{Block-DOMS. Voxels of neighboring blocks in
  $y^{-}$ and $y^{+}$ direction (yellow regions) 
  can be easily located by depth-encoding tables. Voxels of the neighboring block in the $x^{+}$ direction
  can be copied to $Block_{(i, j)}$.}
  \label{fig:block_doms}
  \Description{}  
\end{figure}

\begin{algorithm}[t]
  \caption{Searching Space Confirmation Algorithm}
  \label{alg1}
  
  \KwIn{Block sets: $\mathbb{B} = \{(m,n)| (m,n) \in \mathbb{D}^2\}$, 
  Depth-encoding table set: $\{T_{(m,n)}| (m,n) \in \mathbb{B}\}$, 
  Input Voxels: $\{P_k^{(m,n)}|P_k^{(m,n)} = (x_k, y_k, z_k), (m,n) \in \mathbb{B}, k=1,2,...,N_b\}$,
  
  \qquad \quad Output Voxel: $Q_o^{(i,j)} = (x_o, y_o, z_o), (i,j)\in \mathbb{B}$ }
  
  \KwOut{Search Space $\mathbb{S}_o$} 
  Initialization: $\mathbb{S}_o \gets \emptyset$;
  $\mathbb{S}_o \gets \mathbb{S}_o \cup \{P_k^{(i,j)}| y_k=y_o:y_o+1, z_k=z_o \}$;
  $\mathbb{S}_o \gets \mathbb{S}_o \cup \{P_k^{(i,j)}| y_k=y_o-1:y_o+1, z_k=z_o+1 \}$;

  \# Traverse the three neighbor blocks in y- dir.

  \ForEach{$\Delta_{x}$ in $\{-1, 0, +1\}$}{
    \If{$(x_o + \Delta_{x}, y_o-1, z_o)$ in Block $(i+\Delta_{x},j-1)$}{

      \# Locate via $T_{(i+\Delta(x), j-1)}$

      \ \  $\mathbb{S}_o \gets \mathbb{S}_o \cup \{P_k^{(i+\Delta_{x},j-1)}| y_k=y_o-1, z_k=z_o \}$;
      $\mathbb{S}_o \gets \mathbb{S}_o \cup \{P_k^{(i+\Delta_{x},j-1)}| y_k=y_o-1, z_k=z_o+1 \}$;
    }
  }

  \# Traverse the three neighbor blocks in y+ dir.

  \ForEach{$\Delta_{x}$ in $\{-1, 0, +1\}$}{
    \If{$(x_o + \Delta_{x}, y_o+1, z_o)$ in Block $(i+\Delta_{x},j+1)$}{

      \# Locate via $T_{(i+\Delta(x), j+1)}$

      \ \  $\mathbb{S}_o \gets \mathbb{S}_o \cup \{P_k^{(i+\Delta_{x},j+1)}| y_k=y_o+1, z_k=z_o \}$;
      $\mathbb{S}_o \gets \mathbb{S}_o \cup \{P_k^{(i+\Delta_{x},j+1)}| y_k=y_o+1, z_k=z_o+1 \}$;
    }
  }
  
  \Return{$\mathbb{S}_o$}
  \end{algorithm}

\begin{figure*}[ht]
  \centering
  \includegraphics[width=\textwidth]{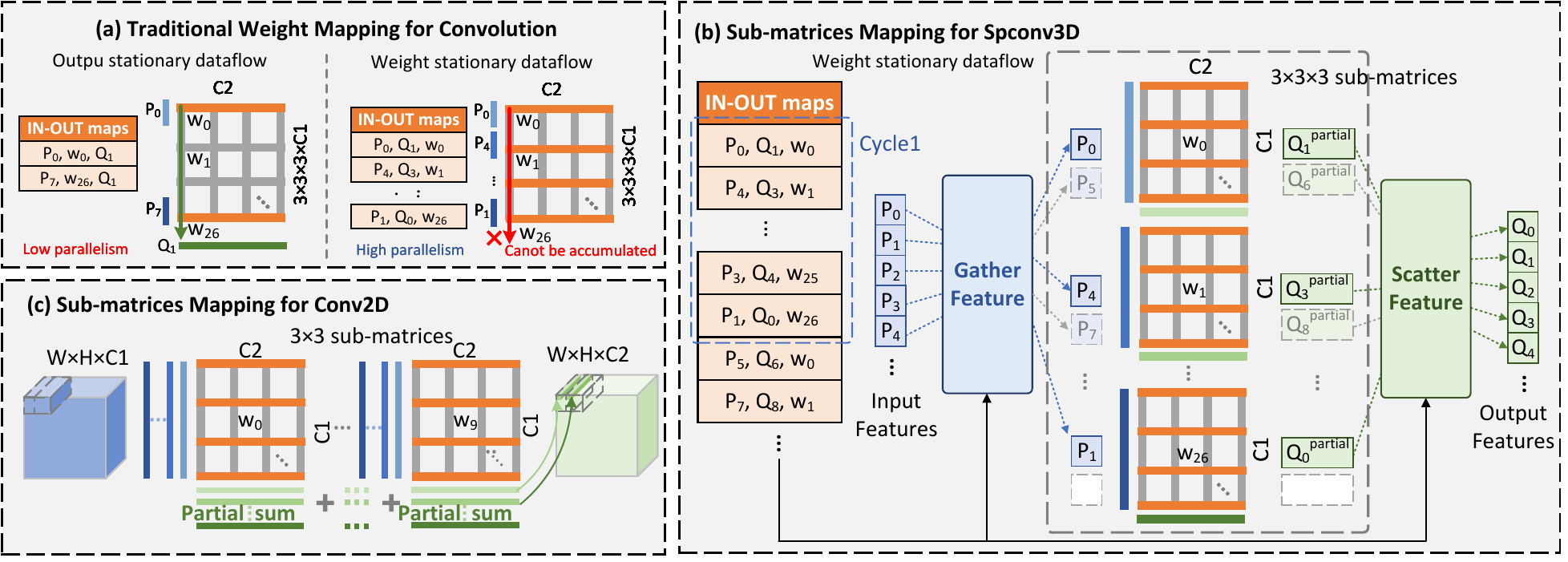}
  \caption{(a) Traditional weight mapping method for convolution 
  (b) Sub-matrices mapping method for Spconv3D (c) Sub-matrices mapping method for Conv2D (Take K=3
  as an example)}
  \label{fig:cimmapping}
  \Description{}  
\end{figure*}

With the assistance of DOMS, our buffer only needs to hold a portion of corresponding
depth. 
However, each dpeth's voxels need to be loaded twice 
(once for the target in previous depth and once
for the target in current depth), resulting in $O(2N)$ memory access.
If the FIFO buffer is large enough to hold voxels of the entire depth, 
the release of the voxels in depth$@z0+1$ could be avoided 
at \textbf{Step\ding{175}}, which can realize $O(N)$ for memory access.

\vspace*{7pt}

{\noindent\itshape D. Blocked Depth-encoding-based Output Major Search}


Furthermore, to achieve stable $O(N)$ memory access volume in various situations, we propose block-DOMS.
This method divides 
the voxel space into 
2D grids to downsize each depth. After data is re-organized at the memory, the 
dominant obstacle lies in how to conduct cross-block search efficiently.
As the example shown in Fig. ~\ref{fig:block_doms}, 
for the output voxel in $Block_{(i, j)}$ (blue region), the adjacent voxels from 
neighbor blocks in
$y^{-} dir.$ and $y^{+} dir.$ (yellow regions) 
can be easily located in memory by depth-encoding tables, 
which must be located at the beginning or end of each depth.
But the adjacent voxels from neighbor block in $x^{+} dir.$ are hard to be located 
and the 
entire depth need to be loaded for further searching.
To address this problem, we copy the adjacent voxels from neighbor block ($Block_{(i+1, j)}$) 
and save them in $Block_{(i, j)}$, 
to avoid cross-block searching in $x^{+} dir.$. 
The $x^{-} dir.$ doesn't need to 
be considered due to symmetry.
Experiments show that the replicated voxels
only account for less than $6\%$ of all voxels, which is negligible.
The detailed Searching Space Confirmation
algorithm is shown in Alg. ~\ref{alg1}.
Each block needs a depth-encoding table. 
Hence, there is a trade-off between table size and searching efficiency.

\subsection{Weight Mapping Strategies for CIM Unit}

CIM is emerging as an efficient paradigm to address the memory wall bottleneck. The weights can be stored in the crossbar-like memory arrays and inputs can activate multiple rows. Thus, the multiply-and-accumulate (MAC) operations can be performed in parallel, which significantly reduces the massive data access.
Previous CIM-based DNN accelerators only focused on
dense Conv2D operations, mainly used in 2D image applications \cite{song2017pipelayer, jiang2020mint, yang2021pimgcn, li202240}.
There are no attempts to accelerate Spconv3D with CIM 
architecture before. In this section, we propose a CIM-based computing unit, which can support 
both Spconv3D and Conv2D operations flexibly.

\vspace*{7pt}

{\noindent\itshape A. Weight Mapping Methods for Spconv3D $\&$ Conv2D}

In traditional weight mapping method for CIM-based accelerators, every channel of the kernel 
is unrolled into a long column and mapped into a column of memory array.
The value in $column-axis$ will be accumulated to get output 
features. As illustrated in Fig. ~\ref{fig:cimmapping} (a), for a typical subm3’s kernel with size $C1 \times K \times K \times K \times C2$,
every channel $(C1 \times K \times K \times K)$ is unrolled into a long column, $C2$ columns in total.
In this mapping method, the input features of each cycle will be multiplied by the weights
and accumulated 
to get the identical feature tensor.
However,
this mapping method doesn't support Spconv3D operations efficiently.
In Spconv3D operations, if we use output stationary dataflow, we will waste lots of CIM computational resources 
due to low parallelism caused by input's sparsity;
if use weight stationary dataflow, the in-out pairs
for each weight correspond to different outputs and cannot be accumulated.
Therefore, we need to design a more efficient weight mapping method for Spconv3D.

Inspired by \cite{peng2019optimizing}, we propose a sub-matrices mapping method for Spconv3D.
The weights at different kernel locations are mapped into 
different sub-matrices. As shown in Fig. ~\ref{fig:cimmapping} (b), for subm3's
kernel with size $C1 \times K \times K \times K \times C2$, 
there will be $K \times K \times K$ sub-matrices,
and each sub-matrix will have a size of $C1 \times C2$.
This method allows each weight to be independently 
controlled for activation or idling, 
providing more flexible support for Spconv3D.
To maximize the utilization of CIM parallel computational resources, we adopts the weight
stationary dataflow. 1) In each cycle, 
the gather unit will gather all features for all
weights of this layer as much as possible according to IN-OUT maps info.
2) Then, the input features will be multiplied with associated weights.
3) Scatter and accumulate the partial sum to corresponding output feature tensor according 
to IN-OUT maps.
To maximize the feature reuse, the input batch of each cycle will be selected based on  
the principle of maximizing overlap with the batch of last cycle.

For Conv2D operations in RPN, because the input feature map has structured tensor format in spatial
locations, we use the same sub-matrices mapping method to maximize the feature
reuse. As shown in Fig. ~\ref{fig:cimmapping} (c),
for a typical Conv2D kernel with size $C1 \times K \times K \times C2$, 
there will be $K \times K$ sub-matrices, and each sub-matrix will have a size of $C1 \times C2$.
In this method, the input feature vectors of the current sub-matrix can be reused for 
next sub-matrix in next 
\begin{figure}[t]
  \centering
  \includegraphics[width=\linewidth]{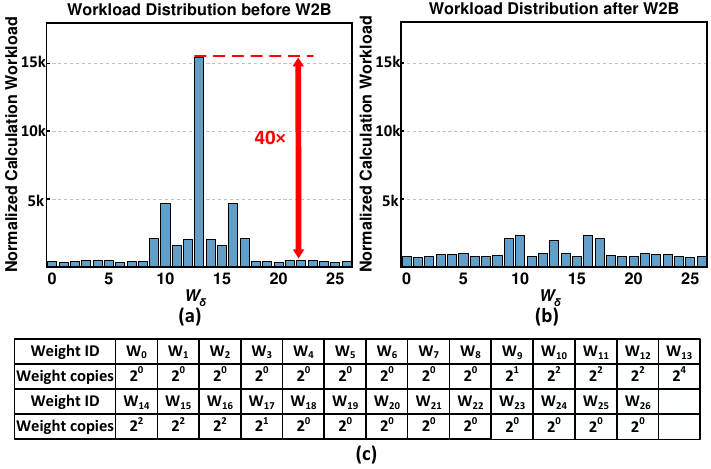}
  \caption{ (a) The workload distribution shows a huge imbalance
  before W2B. (b) The workload imbalance is alleviated after W2B.
  (c) Parameter setting of W2B for the first layer (subm3) of SECOND. }
  \label{fig:w2r}
  \Description{}  
\end{figure}
cycle, when the kernel is sliding on input feature maps.
Hence, it can maximize feature reuse.

\vspace*{7pt}

{\noindent\itshape B. Weight Workload Balanced Method}

In traditional CIM unit, all weights are evenly mapped into memory cells. 
This mapping paradigm is not suitable for computations of irregular point cloud data.
As shown in Fig. ~\ref{fig:w2r} (a), different weights correspond to varying
number of in-out pairs, the workload gap between the central and the
peripheral weights can even more than $40$ times.
Once the computational workload for peripheral weight computing units is completed, 
only the central weight computing unit remains active, which leaves the peripheral 
weights stay idle. This results in a considerable underutilization of 
computational resources.
To balance the computational workload of different weights, 
we replace the evenly distributed weight mapping 
paradigm with the weight workload balanced (W2B) method.
Extra copies are made to the central weights, 
while the edge weights are either not replicated or 
replicated to a lesser extent.
The Fig. ~\ref{fig:w2r} (c) is an optimization example of detailed copying factor parameter
setting for the first layer (subm3) in SECOND.
The weight's workload is evaluated by normalized computational workload, defined by workload/weight copies.
After optimization, the workload distribution becomes more uniform (Fig. ~\ref{fig:w2r} (b)).



\subsection{Overall Architecture}

\begin{figure}[ht]
  \centering
  \includegraphics[width=\linewidth]{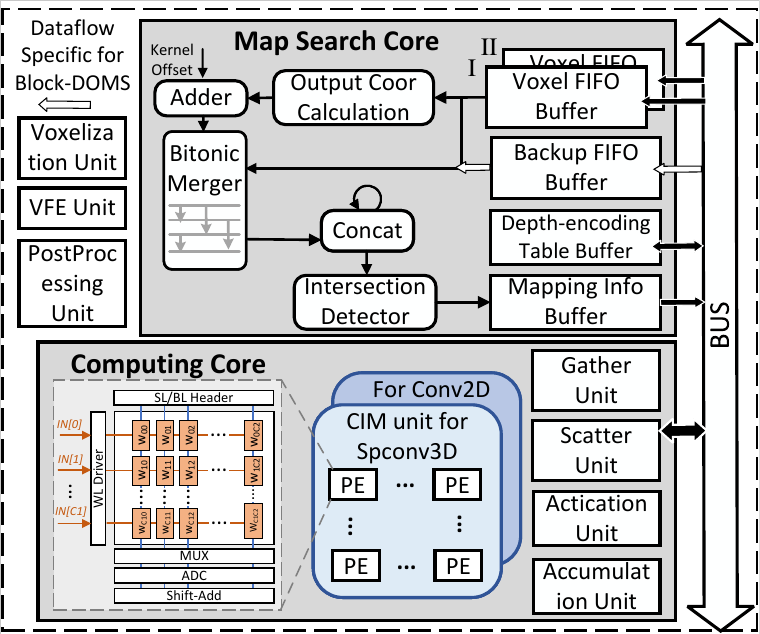}
  \caption{Overview of Voxel-CIM architecture.}
  \label{fig:arch}
  \Description{}  
\end{figure}

As shown in Fig. ~\ref{fig:arch}, Voxel-CIM is mainly composed of two main cores:
map search core and computing core.
The map search core realizes DOMS and block-DOMS operations for efficient kernel
map search. It uses two voxel FIFO buffers (I and II) to store the voxels across two adjacent
depths. The voxel coordinates are loaded to the corresponding buffer row by row
according to the depth-encoding table. Upon
loading, the output voxel coordinates are calculated, 
and its adjacent positions are determined by adding the kernel offset by a parallel adder.
The merge sorter is a bitonic sorter designed for fixed-length sequences.
The intersection detector is
implemented using a comparator that compares three coordinates
\begin{figure}[ht]
  \centering
  \includegraphics[width=\linewidth]{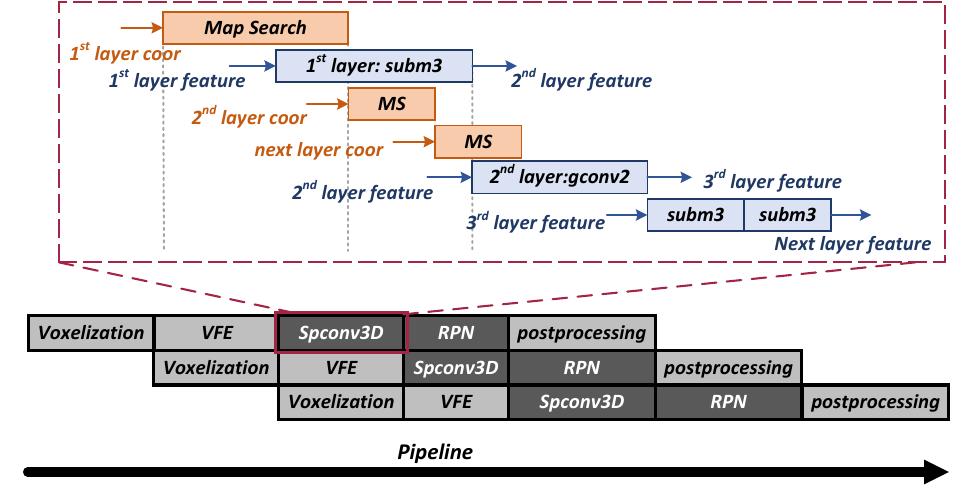}
  \caption{MS-wise pipeline and Compute-wise pipeline.}
  \label{fig:pipeline}
  \Description{}  
\end{figure}
simultaneously in parallel. Valid in-out pairs are temporarily 
stored in the
Mapping Info buffer.
To activate block-DOMS, the backup FIFO buffer is
hired to store the voxel coordinates from neighbor blocks.

In the computing core, a set of CIM units is used to conduct Conv2D and Spconv3D 
by corresponding weight mapping strategies. 
The CIM unit is composed of tiles, where each tile contains 1024 $\times$ 1024 memory cells.
Each cell can store 1 bit information. The tile is divided into different PEs. Each PE contains
all necessary resources to perform MAC operations, such as
MUXs, ADCs, Shift-Adders, etc.
According to sub-matrices mapping method, 
the weight's matrices are mapped to memory in units of PE.

In addition, the voxelization unit is used to partition the point cloud into different voxels in the initial stage,
The VFE unit can support various VFE operations (e.g., dynamic VFE and simple VFE) flexibly.
The post-processing unit is responsible for subsequent processing work, targeting different tasks.

\begin{figure*}[htb]
  \centering
  \includegraphics[width=\textwidth]{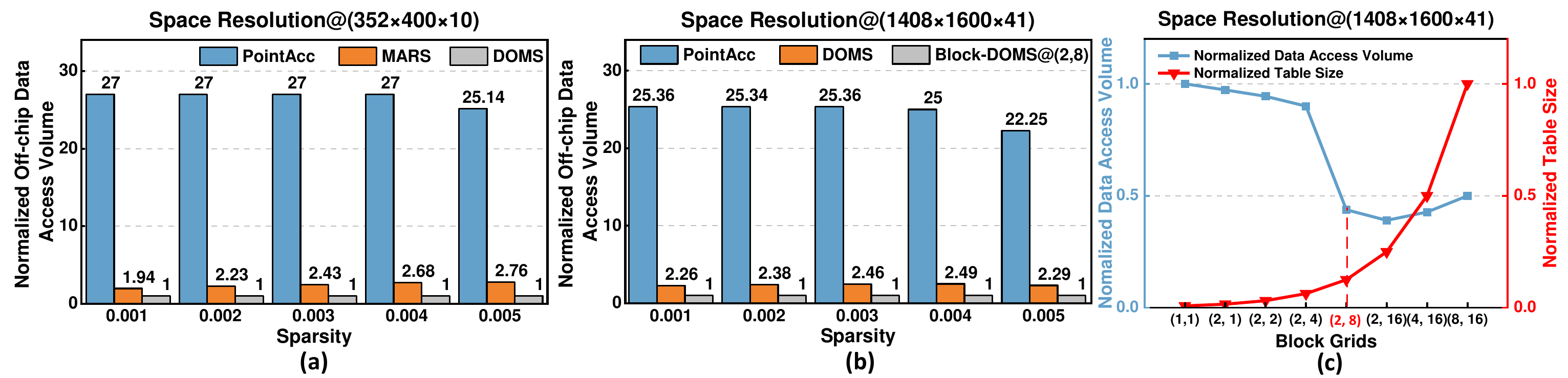}
  \caption{(a) Normalized data access volume comparison in low space resolution.
  (b) Normalized data access volume comparison in high space resolution.
  (c) Trade-off between data access volume and table size.}
  \label{fig:experiment1}
  \Description{}  
\end{figure*}

Fig. ~\ref{fig:pipeline} shows the hybrid pipeline for detection tasks, containing MS-wise pipeline
and compute-wise pipeline.
The Map Search Core follows the MS-wise pipeline:
the map search (MS) for $2^{nd}$ layer does not depend on the results of the 
$1^{st}$ layer's convolution; therefore, as soon as the 
MS for $1^{st}$ layer is complete, the $2^{nd}$ layer's MS can commence immediately.
The computing core follows the compute-wise pipeline:
the $1^{st}$ layer's convolution does not require 
the completion of its entire map search; it can be executed as 
long as there are a sufficient number of in-out pairs.
The $2^{nd}$ layer's compute needs to wait the completion of $1^{st}$ layer.
Two consecutive subm3 layers share common IN-OUT maps, without changing the 
position of voxels, thus the latter subm3 layer doesn't require MS again.

\section{Evaluation}

\vspace*{7pt}

{\noindent\itshape A. Evaluation Setup}

\begin{table}[h]
\centering
\caption{EVALUATION BENCHMARKS}
\label{tab:my_label}
\begin{tabular}{ccccc}
\hline
Application & Dataset & Model & Notation \\ \hline
Detection & KITTI\cite{geiger2012we} & SECOND\cite{yan2018second}  & Det \\
Segmentation & SemanticKITTI\cite{behley2019semantickitti} & MinkUNet\cite{choy20194d} & Seg \\ \hline
\end{tabular}
\end{table}

\textbf{Benchmarks.}
We choose 2 classic voxel-based networks, SECOND \cite{yan2018second} and 
MinkUnet \cite{choy20194d} to evaluate performance 
on detection and segmentation tasks respectively, as shown in Table ~\ref{tab:my_label}.
Experiments are conducted on 2 widely used benchmarks, KITTI and SemanticKITTI.
Like \cite{lyu2023spocta}, all weights of models are quantized to 8 bits. 

\textbf{Hardware Simulation.}
To evaluate the performance of map search methods, we developed a simulator based on Python.
This simulator can generate random voxel data with varying space resolution and sparsity
according to our setting.
The behavior of searching methods will be modeled to analyze the off-chip data access volume and 
sorting times.
The hardware performance of Voxel-CIM architecture is evaluated with NeuroSim framework \cite{peng2020dnn+}. 
The overhead of other units such as Voxelization, VFE etc. are evaluated on Xeon Platinum 8358P CPU.

\textbf{Baseline.}
Several representative hardware platforms are selected as baselines, including four start-of-the-art
Spconv-based accelerators and two powerful Nvidia GPUs. Standard networks, such as SECOND 
and MinkUnet, are implemented on NVIDIA GPU and Xeon CPU to evaluate the overall end-to-end
performance.

\vspace*{7pt}

{\noindent\itshape B. Evaluation Results}

{\itshape 1) Comparison of Map Search Methods.}

To analyze the limitation of the sorter buffer, we set the buffer size to match the length of merge
sorter, that is 64.
Fig. ~\ref{fig:experiment1} shows our simulation results with varying space resolution and sparsity cases.
When the space resolution is low (e.g. $352 \times 400 \times 10$), 
both MARS and DOMS show superiority to PointAcc. But with the sparsity increases,
the data access volume of MARS increases slightly while DOMS shows stable optimal performance
with $O(N)$ access. 
When the space resolution is high (e.g. $1402 \times 1600 \times 41$),
the data access volume of DOMS also increases slightly, nearly to $O(2N)$. 
Accordingly, we employ block-DOMS to optimize the searching 
process. 

To find a ideal block partition factor, we 
analyze the depth-enco-ding table size and data access volume with fixed sparsity@0.005.
As shown in Fig. ~\ref{fig:experiment1}(c), smaller block granularity tends to result in less data access volume, 
yet the size of the table increases accordingly. 
Additionally, when the number of blocks reaches a certain threshold, 
data access volume may increase due to the replicated voxel in $x+ dir.$. 
Given the trade-off between data access volume and table size, 
the most optimal block partition factor is $(2, 8)$ for this specific case. 
As illustrated in Fig. ~\ref{fig:experiment1}(b), it is obvious that the block-DOMS@(2,8)
can maintain $O(N)$ data access volume in various situations.

{\itshape 2) Weight Workload Balanced Method.}

\begin{figure}[h]
  \centering
  \includegraphics[width=\linewidth]{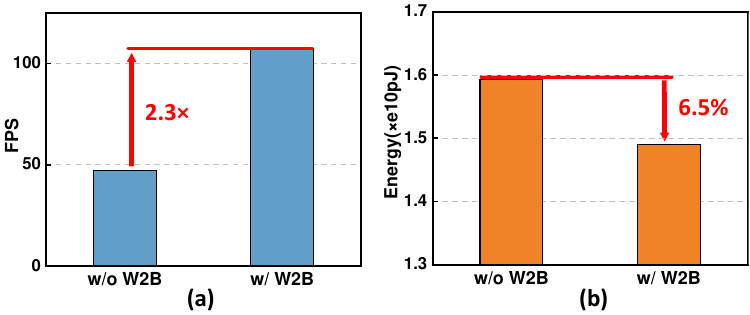}
  \caption{Performance improvement with W2B in terms of FPS and energy consumption for segmentation task.}
  \label{fig:w2r_eva}
  \Description{}  
\end{figure}

\begin{figure}[h]
  \centering
  \includegraphics[width=\linewidth]{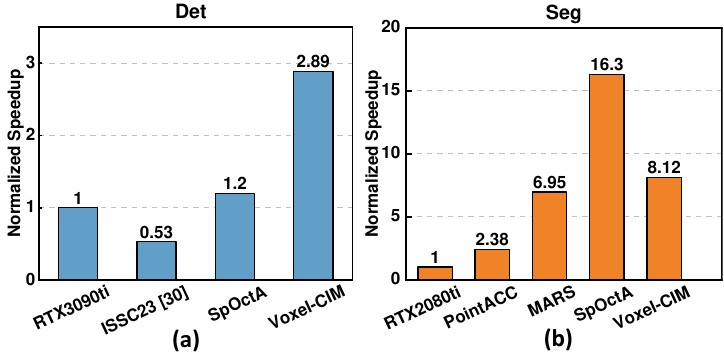}
  \caption{Normalized speedup compared to previous works and powerful GPU in 
  detection and segmentation tasks.}
  \label{fig:overall_eva}
  \Description{}  
\end{figure}

We propose the W2B method to solve the weight workload imbalanced problem in Spconv3D.
The experiments are conducted on MinkUnet for segmentation tasks because 
the segmentation network primarily consists of Spconv3D layers.
The experiment results are shown in Fig. ~\ref{fig:w2r_eva}.
After W2B, the imbalance issue has been mitigated, 
achieving a 2.3$\times$ speedup and 6\% decrease in energy consumption.

{\itshape 3) Comparison with the State-of-the-Art Works}

The implementation details and overall performance are summarized in Table ~\ref{tab:comparison}.
Voxel-CIM is implemented using 22 nm technology and operates at a frequency of 1000MHz.
The CIM component is based on SRAM memory cells.
The Voxel-CIM achieves a peak throughput of 27.8 TOPS.
Compared to other Spconv-based accelerators, the Voxel-CIM achieves 10.8 TOPS/W @ 0.85V, showing 4.5\~{} 7.0$\times$ higher energy efficiency.
As shown in Fig.  ~\ref{fig:overall_eva},
for object detection task in KITTI benchmark, the Voxel-CIM achieves 2.89$\times$ speedup over 
a powerful GPU 3090ti, and 2.4$\times$ speed up over the state-of-the-art works.
For segmentation task in SemanticKITTI benchmark, our accelerator achieves 8.12$\times$ speedup over 
a powerful GPU 2080ti. Although it is slower than the state-of-the-art work in terms of FPS, 
it offers a $4.5\times$ higher energy efficiency.

\begin{table}
  \caption{COMPARISON WITH OTHER WORKS}
  \label{tab:comparison}
  \centering
  \small
  \resizebox{\linewidth}{!}{
  \begin{tabular}{|m{2.5cm}|c|c|c|c|c|c|} 
    \Xhline{1.3pt} 
    \textbf{Chip} &  \textbf{PointACC} \cite{lin2021pointacc} & \textbf{MARS} \cite{yang2023efficient} & \textbf{ISSCC23} \cite{sun202328nm} & \textbf{SpOctA} \cite{lyu2023spocta} & \textbf{Voxel-CIM} \\
    \Xhline{1.3pt} 
    \textbf{Tech (nm)} & 40 & 40 & 28 & 40 & 22 \\
    \hline
    \textbf{Frequency (MHz)} & 1000 & 1000 & 450 & 400 & 1000 \\
    \hline
    \textbf{Buffer (KB)} & 776 & 776 & 176 & 177.4 & 776 \\
    \hline
    \textbf{DRAM Bandwidth} & HBM2 250GB/s & HBM2 250GB/s  & - & DDR4 16GB/s & HBM2 250GB/s \\
    \hline
    \textbf{Peak Throughput (GOPS)} & 8000 & 8000 & 225 & 200 & 27822 \\
    \hline
    \textbf{Peak Energy Efficiency(TOPS/W)} & - & -  & 1.55@0.85V & 2.39@1V & 10.8@0.85v\\
    \hline
    \textbf{Det(fps)}  & - &  - & 19.4 & 44.0 & 106 \\
    \hline
    \textbf{Seg(fps)} & 31.3 & 91.4  & - & 214.4 & 107 \\
    \hline
  \end{tabular}}
\end{table}

\section{Conclusion}
In this paper, we propose Voxel-CIM, a Compute-in-Memory (CIM) based 
accelerator for voxel-based neural network processing.
The novel map search methods,
DOM and block-DOMS, are proposed to reduce off-chip memory access volume,
achieving stable $O(N)$-level data access volume in various extreme 
situations. 
We also design dedicated weight mapping
methods to efficiently support in-memory computing of Spconv3D and Conv2D.
A weight workload balanced mechanism is proposed to
mitigate the issue of workload imbalance and improve computational resources utilization.
Extensive experiments show that 
Voxel-CIM can achieve averagely 4.5\~{}7.0$\times$ higher energy efficiency, 2.4\~{}5.4$\times$ speed up in detection task 
and 1.2\~{}8.1$\times$ 
speed up in segmentation task compared to the state-of-the-art point cloud accelerators 
and powerful GPUs.

\section*{ACKNOWLEDGMENTS}
This work is supported in part by the National Natural Science Foundation of 
China (Grant No. 62304195), Guangzhou-HKUST(GZ) Joint Funding Program 
(No. 2024A03J0541), Guangzhou Municipal Science and Technology Project 
(Municipal Key Laboratory Construction Project, Grant No.2023A03J0013)

\bibliographystyle{IEEEtran}
\bibliography{sample-base}

\end{document}